\documentclass{elsart}
\usepackage{natbib}
\usepackage[dvips]{graphics, graphicx, epsfig}
\usepackage{caption}
\usepackage{amsmath}

\begin{document}
\runauthor{F. Wauters et al.}
\begin{frontmatter}
\title{A GEANT4 Monte-Carlo Simulation Code for precision $\beta$ spectroscopy}

\author[IKS]{F. Wauters},
\author[IKS]{I. Kraev},
\author[NPI]{D. Z\'akouck\'y},
\author[IKS]{M. Beck}\footnote{\textit{Present address}: Westf\"alische Wilhelms-Universit\"at M\"unster, Institut f\"ur Kernphysik, Wilhelm-Klemm-Stra\ss e 9, 48149 M\"unster, Germany},
\author[IKS]{V.V. Golovko}\footnote{\textit{Present address}: Department of Physics, Queen's University, Stirling
Hall, Kingston, ON, Canada, K7L3N6},
\author[IKS]{V.Yu. Kozlov}\footnote{\textit{Present address}: Karlsruher Institut f\"ur Technologie, Karlsruhe, Germany},
\author[IKS]{T. Phalet},
\author[IKS]{M. Tandecki},
\author[IKS]{E. Traykov},
\author[IKS]{S. Van Gorp},
\author[IKS]{N. Severijns \corauthref{cor1}}
\ead {nathal.severijns@fys.kuleuven.be} \corauth[cor1]{Tel.:
+32-16-327265; fax: +32-16-327985.}
\address[IKS]{K.U. Leuven, Instituut voor Kern- en Stralingsfysica, Celestijnenlaan 200D, B-3001 Leuven, Belgium}
\address[NPI]{Nuclear Physics Institute, ASCR, 250 68 \v{R}e\v{z}, Czech Republic}

\begin{abstract}
The measurement of the $\beta$ asymmetry parameter in nuclear $\beta$ decay is a potentially very sensitive tool to search for non $V-A$ components in the charge-changing weak interaction. To reach the required precision (percent level) all effects that modify the emission pattern of the $\beta$ radiation, i.e. the geometry of the setup, the effect of the magnetic field on the trajectories of $\beta$ particles as well as (back)scattering in the source, on the sample holder and on the detector, have to be correctly taken into account in the analysis of the data. A thorough study of these effects and a new method based on detailed GEANT4 Monte-Carlo simulations that was developed for this purpose is presented here. The code was developed for $\beta$ asymmetry measurements by means of the Low Temperature Nuclear Orientation (LTNO) method, but can in principle be generalized to other experimental setups using other polarization techniques.   \end{abstract}


\end{frontmatter}

\section{Introduction and motivation}

In nuclear $\beta$ decay the search for physics beyond the Standard Electroweak Model is a key issue in a series of precision experiments that are carried out or being prepared now (see e.g. \cite{Adelberger:99}-\cite{Wauters2009}).
%\cite{Adelberger:99, Crane:2001, Gorelov:2005, Kozlov:2006a, Kozlov:2006b, Melconian07, Vetter08, Flechard08, Pitcairn09, Wauters2009}).
One of the powerful techniques to probe the structure of the Standard Model is a measurement of the angular distribution of $\beta$ radiation with the Low Temperature Nuclear Orientation (LTNO) method.
One of the best examples of this is the famous experiment carried out by C.S. Wu et al. in 1957 \cite{Wu:1957} and later repeated with better precision by Chirovsky et al. \cite{Chirovsky:1980}. The observed angular distribution was found to be in agreement with the expected
$
1 + AP\left( {{v \mathord{\left/
 {\vphantom {v c}} \right.
 \kern-\nulldelimiterspace} c}} \right)\cos \left( \theta  \right)
$ dependence from the $V$-$A$ theory of the weak interaction (the
explanation of the parameters in this formula is given below).
Such a measurement of the angular distribution is a potentially
very sensitive tool to search for non $V-A$ (i.e. tensor and
scalar) components in the weak interaction. However, present upper limits
for such contributions (which are at the level of 8\% (95~\% C.L.) in
the amplitudes \cite{Severijns2006}) require the angular distribution coefficient
$\tilde{A}$ (see below) to be determined with a precision better
than about 1~\%. In the past we have performed already several measurements of the
angular distribution of $\beta$ radiation with the LTNO method
\cite{Severijns1989}, \cite{Schuurmans2002}, \cite{Severijns2005}.
However, in order to reach the 1~\% precision level the method had
to be further optimized. Therefore, a new LTNO setup was
recently installed at the Katholieke Universiteit Leuven \cite{Kraev:2005},
while a GEANT4 based Monte Carlo code for this type of measurements
was developed as well. These Monte-Carlo simulations turned out to be
a crucial factor, providing a detailed understanding of systematic effects
such as e.g. scattering and magnetic field related effects.
The method presented here is primarily meant to search
for a tensor current admixture in the $V$-$A$ structure
of the weak interaction. It is also being used for the analysis and interpretation of
measurements performed at the ISOLDE-facility at CERN \cite{Severijns2004}
and can in principle be generalized to other experimental setups using
different polarization techniques as well.

The measured angular distribution of $\beta$ radiation
emitted by oriented nuclei can be expressed as \cite{Kraev:2005}:
\begin{equation}
W\left( \theta  \right) =
\frac{{N\left( \theta  \right)_{pol} }}{{N\left( \theta  \right)_{unpol} }}
 = 1 + {v \over c}\tilde{A} \cdot P \cdot
Q_1 \cdot \cos \theta \label{W}
\end{equation}
\noindent Here $N\left(\theta\right)_{pol}$ ($N\left(\theta\right)_{unpol}$) is
the count rate in the detector at an angle $\theta$ with respect
to the nuclear orientation axis for polarized (unpolarized) nuclei,
${v/c}$ is the
$\beta$ particle velocity relative to the speed of light, $P$ is
the degree of nuclear polarization and $\tilde{A}$ is the angular
distribution coefficient that is sensitive to the possible presence
of tensor currents. The modification of the angular
distribution pattern of the $\beta$ radiation due to the
finite sizes of the detector and the source, (back)scattering of
the $\beta$ particles and the presence of a magnetic field is taken into account
by the so-called solid angle correction factor $Q_1$.
However, as will be pointed out below, the general approach to
calculate this $Q_1$ factor \cite{Stone:1986} is not applicable anymore
when the influence of the disturbing effects becomes significant
and/or high precision is required. Therefore, a new method based on
Monte-Carlo simulations with the GEANT4 toolkit \cite{geant:2003} was
developed to correct the anglular distribution for the above
mentioned effects.
As the first measurements of the $\beta$ asymmetry
parameter with our new LTNO setup were performed with $^{60}$Co,
most of the simulations were done for this isotope. The experimental
setup was described previously already in detail \cite{Kraev:2005}.
Only a short description will therefore be given here. This paper
will deal mainly with the GEANT4 simulations. The performance of
GEANT4 for electron backscattering and electron energy deposition
in the energy range from 0.040~MeV to 1.0~MeV was recently investigated
by Martin et al. \cite{Martin:2003, Martin:2006} and by Kadri et al.
\cite{Kadri:2007}. Good agreement with experimental data at the
level of a few percent was obtained in all cases. Recently, GEANT4
simulations were also performed for scintillation detectors, leading
to similar conclusions \cite{Golovko:2008}.

\section{Experimental setup and solid angle correction factor}

\begin{figure}[ht]
\centering
\includegraphics[scale=0.5]{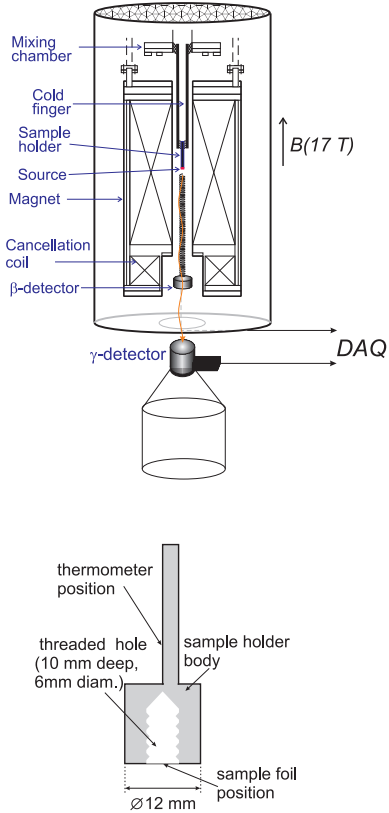}
\caption {Schematic view of the lower part of the experimental
setup, showing also the detailed geometry of the sample holder
and the position of the source foil (containing the $^{60}$Co
and $^{57}$Co activities) and of the $^{54}$Mn thermometer.
The magnetic field is pointing vertically.}
\label{fig:sketch}
\end{figure}

In order to orient nuclei the Brute Force LTNO method is used here. The setup consists of a $^3$He-$^4$He dilution refrigerator, which permits to cool samples to a temperature of about 5-10 mK, in combination with a superconducting magnet. A schematic view of the lower part of the setup is presented in Fig.~\ref{fig:sketch}. The magnet was manufactured by Oxford Instruments Ltd. and can produce a field up to 17 T. $\beta$ particles emitted by the radioactive nuclei are focused by the strong magnetic field and registered by the particle detector. The choice of particle detector is determined by its ability to work both at temperatures close to liquid He temperature and in a strong magnetic field. Si PIN photodiodes produced by Hamamatsu Photonics (type S3590-06) with a thickness of 500 $\mu$m and an active
surface of 9$\times$9 mm$^{2}$ have been tested and showed good behavior under such conditions \cite{Wauters:2009}. Detectors of this thickness are able to fully stop electrons with energies up to about 350 keV.
Note that the 500 $\mu$m thickness is thick enough for the isotopes of interest for this type of measurements (e.g $^{60}$Co, $^{95}$Nb, $^{133}$Xe, etc.) and at the same time thin enough to have low sensitivity to $\gamma$ quanta. The source-to-detector distance is about 20 cm and with the central field of 17 T the magnetic field at the site of the Si detector is about 0.8 T.

The sample for measurements of the $^{60}$Co $\beta$ asymmetry parameter was prepared by diffusing $^{60}$Co activity into a copper foil with a thickness of 20 $\mu$m and a purity of 99.99+~\%. A special temperature regime was used to limit the depth of diffusion to a few microns \cite{Kraev2006}. This foil was then soldered onto the oxygen-free copper sample holder which is, via the cold finger, in thermal contact with the $^{3}$He-$^{4}$He mixing chamber where the lowest temperature of the refrigerator is reached. Geometrically the foil is positioned in
the center of the magnetic field and perpendicular to the field direction (Fig.~\ref{fig:sketch}).

To determine temperatures in the millikelvin region a nuclear orientation thermometer (e.g. $^{57}$Co(\underline{Fe}) or $^{54}$Mn(\underline{Ni})) is soldered on to the sample holder. The temperature is then deduced from the degree of anisotropy observed for the $\gamma$ rays of this thermometer. The $\gamma$ rays are registered by a high-purity Ge $\gamma$ detector installed outside the refrigerator at a distance of about 30 cm from the center of the magnetic field (see Fig.~\ref{fig:sketch}).

From the above given details of the experimental setup it can be seen that the following effects can modify the angular distribution pattern of the $\beta$ radiation:
\begin{itemize}
\item[-] scattering of the $\beta$ particles in the host foil material;
\item[-] backscattering on the layer of solder;
\item[-] backscattering on the sample holder;
\item[-] backscattering on the detector;
\item[-] magnetic field influence on the $\beta$ particle propagation;
\item[-] possible Compton background from $\gamma$ rays accompanying the $\beta$ decay and/or $\gamma$ rays emitted by the nuclear orientation thermometer.
\end{itemize}
As was already mentioned, the distortion of the emitted angular distribution of the $\beta$ radiation is taken into account by introducing the solid angle correction factor $Q_1$ (see Eq. \ref{W}). A general formalism to calculate the $Q_1$ factor is described in Ref.~\cite{Stone:1986}. For an axially symmetric detection system this $Q_1$ factor is defined as
\begin{equation}
Q_1 = \frac{{\int {\sin \left( \theta  \right)} \cos \left( \theta  \right)\varepsilon \left( \theta  \right)d\theta }}{{\int {\sin \left( \theta  \right)\varepsilon \left( \theta  \right)} }}
\label{Q1}
\end{equation}
where $\theta$ is the angle of the $\beta$ particle emission direction with
respect to the orientation axis (which coincides in this case with the detector axis) and $\varepsilon\left( \theta  \right)$ is the detector response function which describes the efficiency of the detector for radiation incident at angle $\theta$. The problem of determining $Q_1$ thus reduces essentially to the knowledge of the detector response function for the $\beta$ radiation. However in the presence of (back)scattering and a strong magnetic field this approach is not valid for precision measurements. A different and more inclusive method for correcting the angular distribution function was therefore developed.

The new method consists of a Monte-Carlo simulation of the complete experiment. The output of this provides the $\beta$ spectrum registered in the $\beta$ particle detector for both isotropic emission of the $\beta$ radiation (i.e. no nuclear polarization) and anisotropic emission (i.e. when the sample is polarized). The angular distribution of the emitted $\beta$ radiation is calculated for exactly the same degree of nuclear polarization $P$ (see Eq.~\ref{W}) as obtained in the experiment (which can be deduced from the observed anisotropies of the $^{60}$Co $\gamma$ rays) and assuming the Standard Model value for the $\beta$ asymmetry parameter, $\tilde A^{SM}$. The simulated isotropic and anisotropic spectra are then analyzed in exactly the same way as the experimental spectra and the $\beta$ asymmetry parameter $\tilde A$ is extracted using the following relation:
\begin{equation}
\frac{{\tilde A_{\exp } }}{{\tilde A_{SM} }} = \frac{{1 - W\left( \theta  \right)^{exp } }}{{1 - W\left( \theta  \right)^{sim} }},
\label{eq:ratio}
\end{equation}
where $W\left( \theta  \right)^{exp }$ and $W\left( \theta  \right)^{sim}$ are the anisotropy functions (Eq.~\ref{W}) obtained experimentally and from the simulations, respectively.

To perform these simulations the GEANT4 toolkit \cite{geant:2003}, which was designed for the simulation of the passage of particles through matter, was used. The development of the simulation code was performed in several steps. Firstly, a simple simulation routine to calculate electron backscattering coefficients was developed. With this routine the influence of the different GEANT4 parameters, i.e. the cut-for-secondaries ($CFS$) parameter and the so-called $f_r$ parameter, on the calculated backscattering fractions was studied. The $CFS$ parameter determines the production threshold for secondary particles, while the $f_r$ parameter limits the step size for tracking of electrons at the boundary between two materials. These parameters were then fixed at the values that provide the best match with the reference data that were obtained from the literature.
Secondly,  $^{60}$Co spectra were measured in a simple experimental setup and compared to simulations. The same setup was thereafter also used to study the influence of the source backing by adding copper foils of different thickness behind the source. Finally $^{60}$Co $\beta$ spectra were measured in strong magnetic fields inside the dilution refrigerator and again compared to simulations. All these steps are discussed in detail in the following sections.

It is important to note that when this work was started the latest available version of GEANT4 was GEANT4.7.1. During the writing of this paper a new version was released, i.e. GEANT4.8.1. Most of the simulations were then redone and compared to the simulations performed with the earlier version. Although some small deviations between the two results were found, there were no major discrepancies that would change the conclusions of this work. Since the comparison of two versions of GEANT4 is not the aim of this paper this is not further discussed. For all GEANT4 simulations the low-energy electromagnetic processes package was used \cite{webref:lowenergy}.

\section{GEANT simulations of $^{60}$Co $\beta$ spectrum.}

\subsection{Electrons backscattering simulations}

One of the factors that may significantly modify the electron angular distribution pattern is backscattering of the electrons on the host foil material into which the $^{60}$Co ions are diffused, as well as on the layer of solder, on the sample holder and on the detector. Therefore, one of the major requirements for a good simulation program is a good description of the backscattering of electrons.
The development of the program was thus started with simulations to reproduce data on electron backscattering coefficients already available in the literature. Since in the measurements a Si PIN-photodiode was used as a $\beta$ particle detector, backscattering on silicon was of particular interest. To obtain good agreement between our simulations and experimental results available in the literature two parameters in GEANT4 were tuned: the $CFS$ parameter and the $f_r$-parameter, that were defined above. The electron backscattering fractions obtained for different values of these two parameters are listed in Table \ref{tab:tuning}. The simulations were performed for normal incidence of electrons with energy $E_e$ = 250 keV. The corresponding backscattering fraction calculated with the formulas given in Ref. \cite{Tabata1971} is 12.9~\%. Comparing the results of our simulations with this value it may be noted that the value for the $f_r$ parameter should not exceed 0.02. From the practical point of view, however, also the simulation time should be taken into account. Indeed, it was found that the simulation time increases with decreasing values for the $f_r$ and $CFS$ parameters, with the influence of the first one being largest. In view of this the $f_r$ parameter was then fixed at the value of $f_r=0.02$.

In choosing the optimal $CFS$ value for $f_r=0.02$ one has more freedom. Indeed, the calculated backscattered fraction of 12.9~\% is determined with an absolute precision of only about a few percent (see Ref.~\cite{Tabata1971}). Within this precision all $CFS$ values in Table~\ref{tab:tuning} give reasonable results for the case with $f_r=0.02$. The $CFS$ value of 10 $\mu$m was then preferred as it also corresponds to a reasonable simulation time. This arbitrariness in the choice of the $CFS$ parameter value might cause a systematic effect in the determination of the $\beta$ asymmetry parameter and the corresponding systematic error on the final result of an experiment will have to be determined.

The optimization of the values for the $CFS$ and $f_r$ parameters described here was started with GEANT4.7.1. In that version of the toolkit the default value of the $f_r$ parameter was equal to 0.2. However, as becomes clear from Table~\ref{tab:tuning} the adjustment of this parameter is of great importance. In the GEANT4.8.1 release the default value of $f_r$ was set to 0.02 which corresponds to the optimal value we also found already in our initial simulations with GEANT4.7.1, as well as with the value recommended in Ref.~\cite{Kadri:2007}.

In Table \ref{tab:backscattering} our simulation results for the backscattering fraction for two different energies of incident electrons (normal incidence) are compared to results obtained by other authors. For these simulations the $f_r$ parameter and the $CFS$ value were set to 0.02 and 10 $\mu$m, respectively. Our results are compared with the experimental data and simulation results of \cite{Berger1969} and with the semi-empirical calculations of Ref.~\cite{Tabata1971}. Good agreement is observed between the results of this work and results obtained by these other authors.

Finally, our simulations showed that the simulated backscattering fractions of electrons for large angles of incidence (i.e. above 30$^\circ$ with respect to the normal of the detector plane)
are larger than the backscattering fractions calculated using the equations in Ref. \cite{Tabata1971} (for the energy dependence) and Ref. \cite{Kanaya1978} (for the angular dependence) by a factor
of approximately 1.2.

\begin{table}[!h]
\begin{center}
\caption
{Total backscattering fraction (in percent) for different values of the cut-for-secondaries $CFS$ and the $f_r$ parameter for electrons with kinetic energy $E_e = 250$ keV at normal incidence. GEANT4.8.1 with the low-energy electromagnetic package was used.}\smallskip
\label{tab:tuning}
\begin{tabular} {l c c c c}
\hline\hline
&&$f_r$&&\\
$CFS$, $\mu$m&0.2&0.02$^1$&0.002\\
\hline
10&8.70(2)&13.20(3)&13.20(4)\\
5&9.10(2)&12.60(3)&13.50(4)\\
1&10.90(2)&12.50(3)&14.30(4)\smallskip\\
\hline\hline
\end{tabular}\\
\end{center}
\ \ \ \ \ \scriptsize $^1$ default GEANT4.8.1 value
\end{table}

\begin{table}[!h]
\begin{center}
\caption
{Total backscattering fraction (in percent)  of primary electrons with different energies $E_e$ for normal incidence on to silicon.}\smallskip
\label{tab:backscattering}
\begin{tabular} {l| c c c c}
\hline\hline
$E_e$, keV&This work&Tabata$^1$&\multicolumn{2}{c}{Berger$^2$}\\
&(simul.)&&(simul.)&(expt.)\\
\hline
250&13.2&12.9&13.2&14(1)\\
500&11.6&11.4&12.3&14(1)\\
\hline\hline
\end{tabular}\\
\end{center}
\hspace{15pt}\scriptsize $^1$ values calculated with the semi-empirical relation in Ref. \cite{Tabata1971};\\
\hspace*{12pt} \scriptsize $^2$ fraction of events without total energy absorption (see Ref.~\cite{Berger1969} for details).\\
\end{table}

\section {Simplified geometry for the $^{60}$Co spectrum measurement and simulation}

\begin{figure}[ht]
\centering
\includegraphics[scale=0.3]{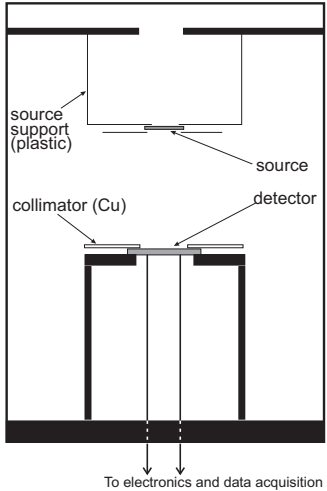}
\caption {Schematic view of the test setup for the $^{60}$Co $\beta$ spectrum measurements.} \label{fig:baby_fridge}
\end{figure}

The next step in the development of the simulation code was to reproduce the $^{60}$Co spectrum measured in a simple geometry and with the same detector as will be used in real measurements. The setup installed for this purpose is shown in Fig. \ref{fig:baby_fridge}.

\subsection {Source preparation and detector configuration}

The source was prepared by drying the $^{60}$Co activity on a 8 $\mu$m thick  mylar foil which  was then covered by another mylar foil of the same thickness to prevent contamination of surrounding materials. The diameter of the $^{60}$Co spot was about 2 mm. The source was attached to a plastic support structure with a 10 mm diameter hole behind the activity spot to avoid backscattering. This source was considered to be a ``scattering free'' source in all later measurements, but the two 8 $\mu$m thick mylar foils were of course included in the simulations.
A 500 $\mu$m thick Si PIN photodiode with a sensitive area of 81 mm$^2$ was used as $\beta$ particle detector. It was mounted on an Al support structure and covered by a 1 mm thick copper collimator with 8 mm diameter hole. The detector and source were axially aligned with a distance of 50 mm between them. They were placed in a vacuum chamber and all measurements were performed at a pressure below 10$^{-4}$ torr.

In all GEANT4 simulations presented in this work the geometry of the setup was included in detail. Special care was taken to describe the particle detector geometry as precisely as possible. For that purpose HAMAMATSU Photonics has provided detailed information on the detector's design (depletion thickness, front and rear dead layers composition and thickness) which was then implemented in the simulation code.

For comparison with the experimental data the simulated spectra were rescaled relative to the experimental ones. The scaling factor was calculated by normalizing the integral of the simulated spectrum with that of the experimental spectrum in the energy region from 150 keV to 300 keV. The same energy region was later used in the analysis of the anisotropy of the $^{60}$Co $\beta$ radiation. The reduced $\chi^2$ value for that energy region was calculated taking into account the finite statistics of the simulations, which is less than that of the experimental data for all simulations shown in this work.

\subsection {Compton background}

\begin{figure}[ht]
\centering
\includegraphics[width=0.6\textwidth]{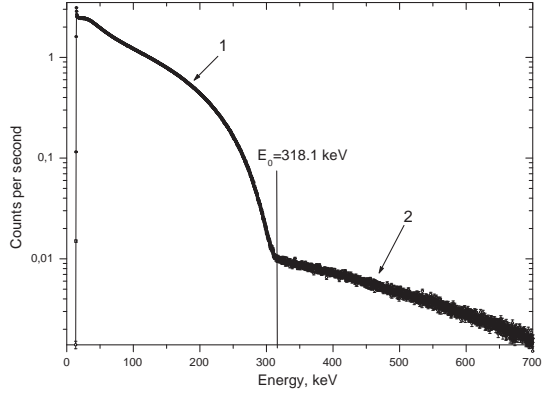}
\caption {$^{60}$Co $\beta$ spectrum obtained with the ''scattering free'' source and the Si PIN photodiode in the test setup. The regions of the spectrum below and above the endpoint energy of $^{60}$Co (i.e. $E_0$ = 318.1 keV) are labeled as 1 and 2 respectively. More details are given in the text.} \label{fig:beta_spectrum}
\end{figure}

\begin{figure}[h]
\centering
\includegraphics[width=0.6\textwidth]{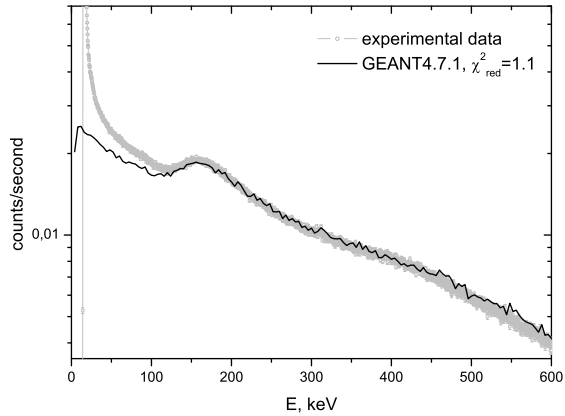}
\caption {Comparison of the experimental and simulated Compton background from $^{60}$Co. Grey dotted line: experimental data (the width of the line is due to the 1$\sigma$ statistical error bar); black solid line: GEANT4.7.1 simulations. The reduced $\chi^2$ for the simulations when compared to the data in the energy range 150 - 300 keV is also indicated.} \label{fig:Compton}
\end{figure}

The $\beta$ spectrum measured with this simple setup is shown in Fig.~\ref{fig:beta_spectrum}.
The tail that extends above the endpoint energy of $^{60}$Co is mainly due to Compton-scattered electrons from the 1173 keV and 1332 keV $\gamma$ lines that accompany the $^{60}$Co $\beta$ decay. The observed $\beta$ spectrum thus contains electrons from both $\beta$ decays and Compton events.
To ensure that the tail is indeed due to Compton scattering of the 1173 keV and 1332 keV $\gamma$ lines a separate measurement, with the detector now covered by a 0.4~mm thick copper plate in which all $^{60}$Co $\beta$ particles are stopped, has been performed. The result is shown in Fig.~\ref{fig:Compton}. The grey line represents the experimental data while the solid line shows the result of the simulations. The detector energy resolution of 7 keV, as determined from measurements with a $^{207}$Bi conversion electron source, was included in these simulations.

As can be seen from Fig.~\ref{fig:Compton}, good agreement with experiment was obtained in the energy region of interest (i.e. 150 - 300 keV). Note that at the lowest energies the experimental spectrum is disturbed by the tail of the electronic noise, which is not included in the simulations. Further, the lowest energy part of the spectrum is dominated by (back)scattered and partially absorbed (Compton) electrons. It might be that GEANT4 is not perfectly describing these processes at such 'low' energies. It is to be noted here that the Compton tail is a smooth function of energy over the entire energy region of interest. This is very important for the analysis of the anisotropy of the $^{60}$Co $\beta$ radiation since the background under the $\beta$ spectrum can cause some uncertainty in the determination of the absolute value of the $\beta$ anisotropy and thus in the determination of the $\beta$ asymmetry parameter $\tilde{A}$.

\subsection{$\beta$ spectrum}

\begin{figure}[ht]
\centering
\includegraphics[width=0.6\textwidth]{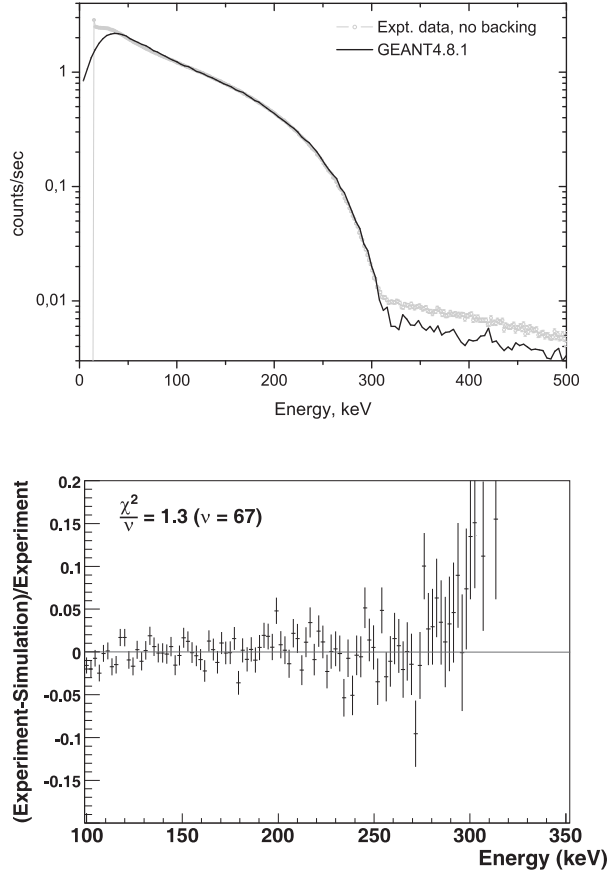}
\caption {(a) Comparison of the experimental $\beta$ spectrum of $^{60}$Co of Fig.~\ref{fig:beta_spectrum} with the simulated one. Grey dotted line: experimental data (error bars are smaller than the size of the data points). Black solid line: GEANT4.8.1 simulations with $f_r = 0.02$ and $CFS$ = 10 $\mu$m. (b) Relative difference between simulated and experimental spectra. In the region of interest the difference is within 2 to 3~\%. It is larger near the $\beta$ spectrum endpoint where the relative contribution of Compton background is more important. The value $\chi^2/\nu$ = 1.3 for 67 degrees of freedom (which is within the 95\% confidence level) refers to the energy region of interest, i.e. from 150 keV to 300 keV.} \label{fig:BabyFr_Comparison_Difference.eps}
\end{figure}

The next step after simulating the Compton background was to reproduce the full $\beta$ spectrum of $^{60}$Co shown in Fig. \ref{fig:beta_spectrum}. The result is presented in Fig. \ref{fig:BabyFr_Comparison_Difference.eps}.
Possible reasons for the discrepancy in the region above 300 keV are discussed below.

\subsection{Effect of backscattering}

\begin{figure}[ht]
\begin{center}
$\begin{array}{c@{\hspace{-0.5cm}}c}
\epsfxsize=7.5cm
\epsffile{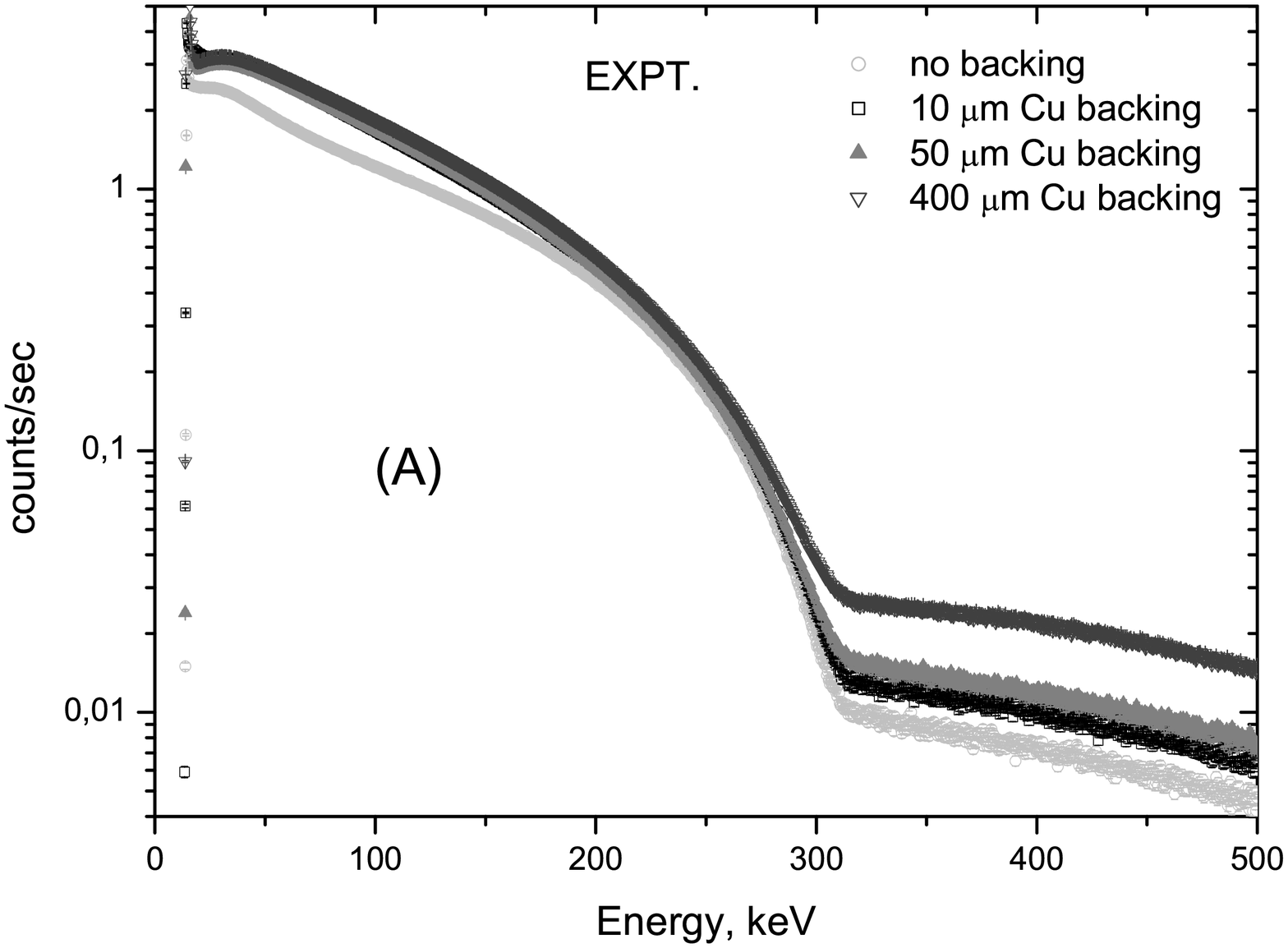}&
\epsfxsize=7.5cm
\epsffile{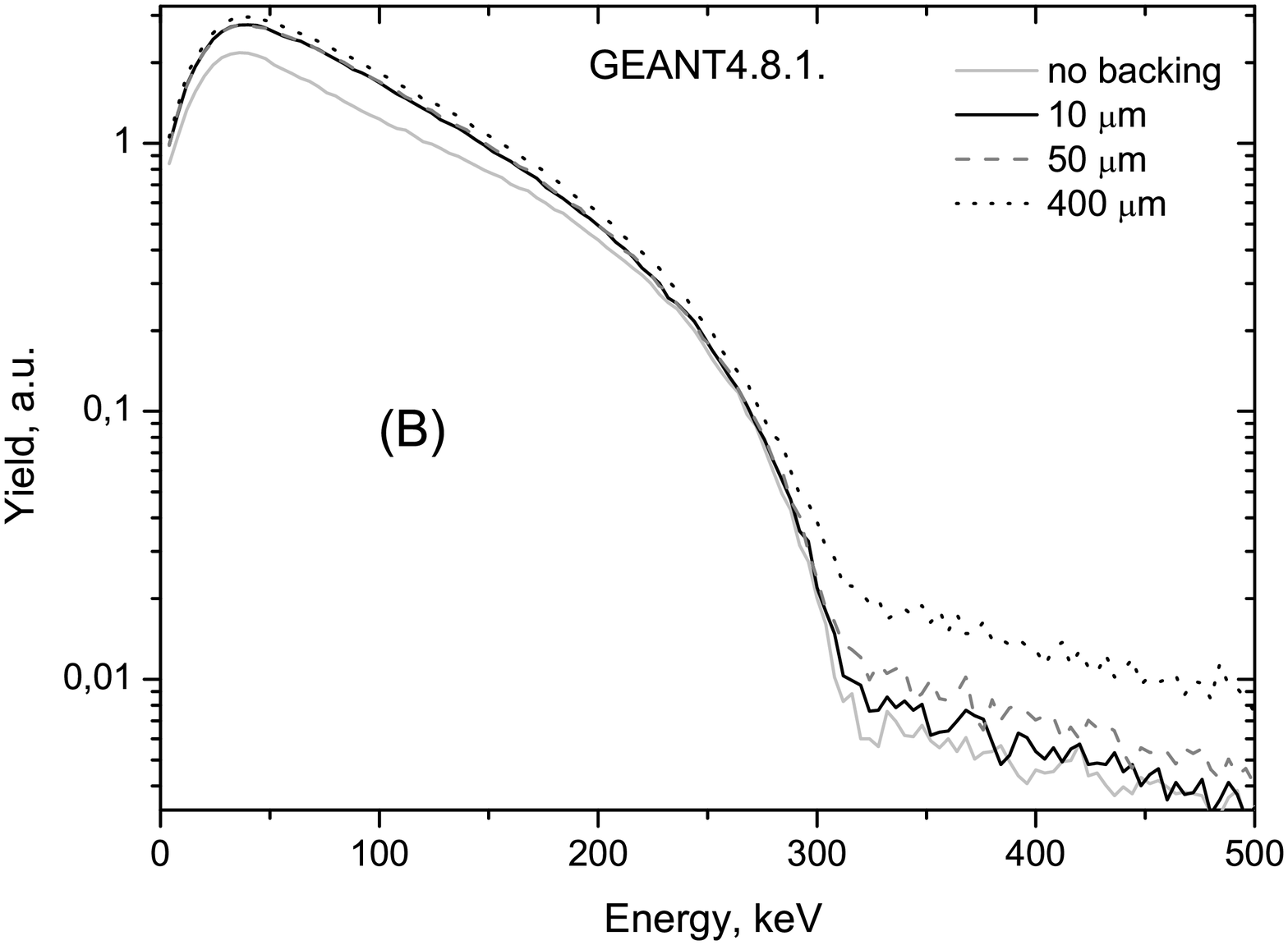}
\\[-0.5cm]
\epsfxsize=7.5cm
\epsffile{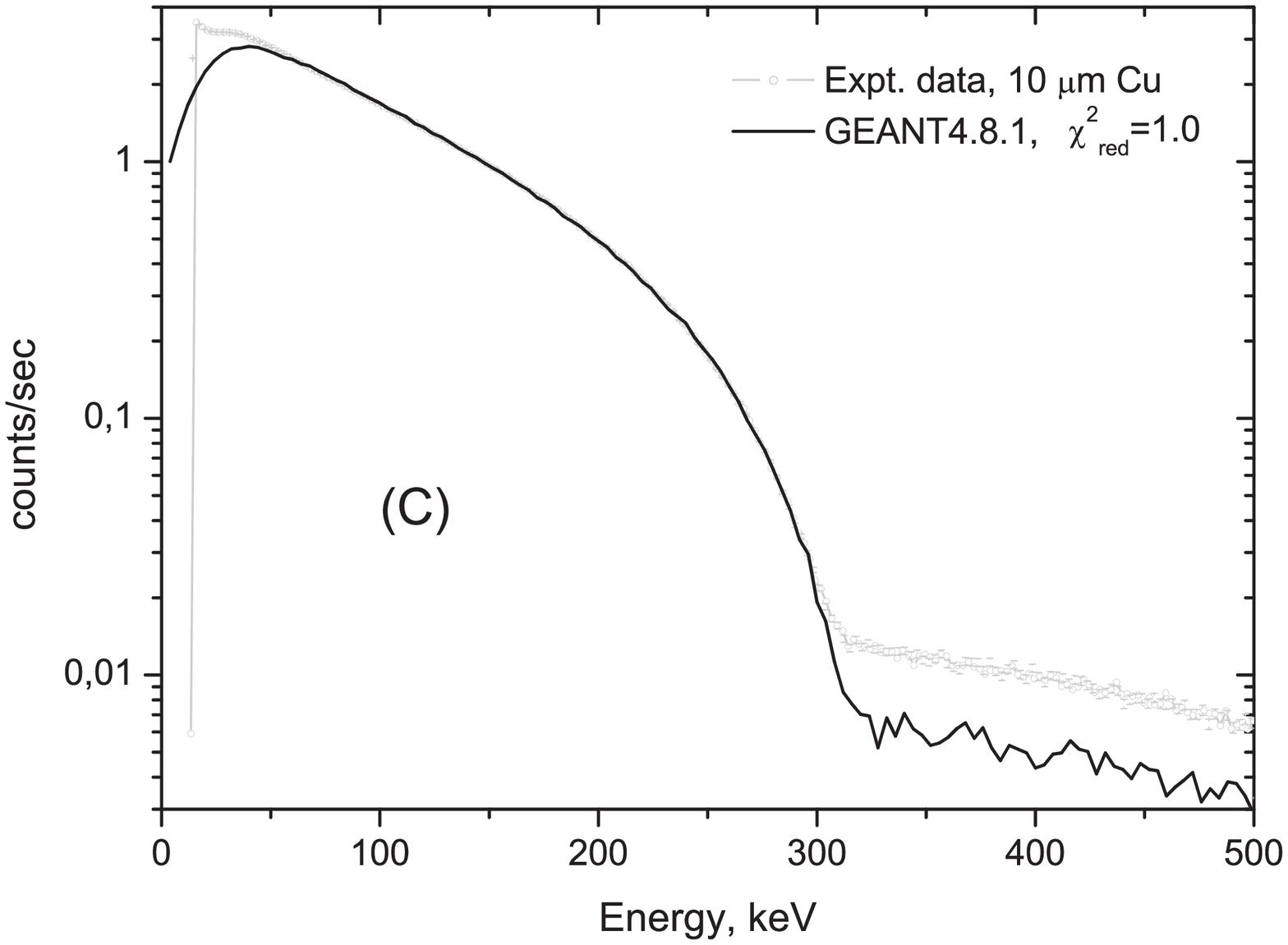}&
\epsfxsize=7.5cm
\epsffile{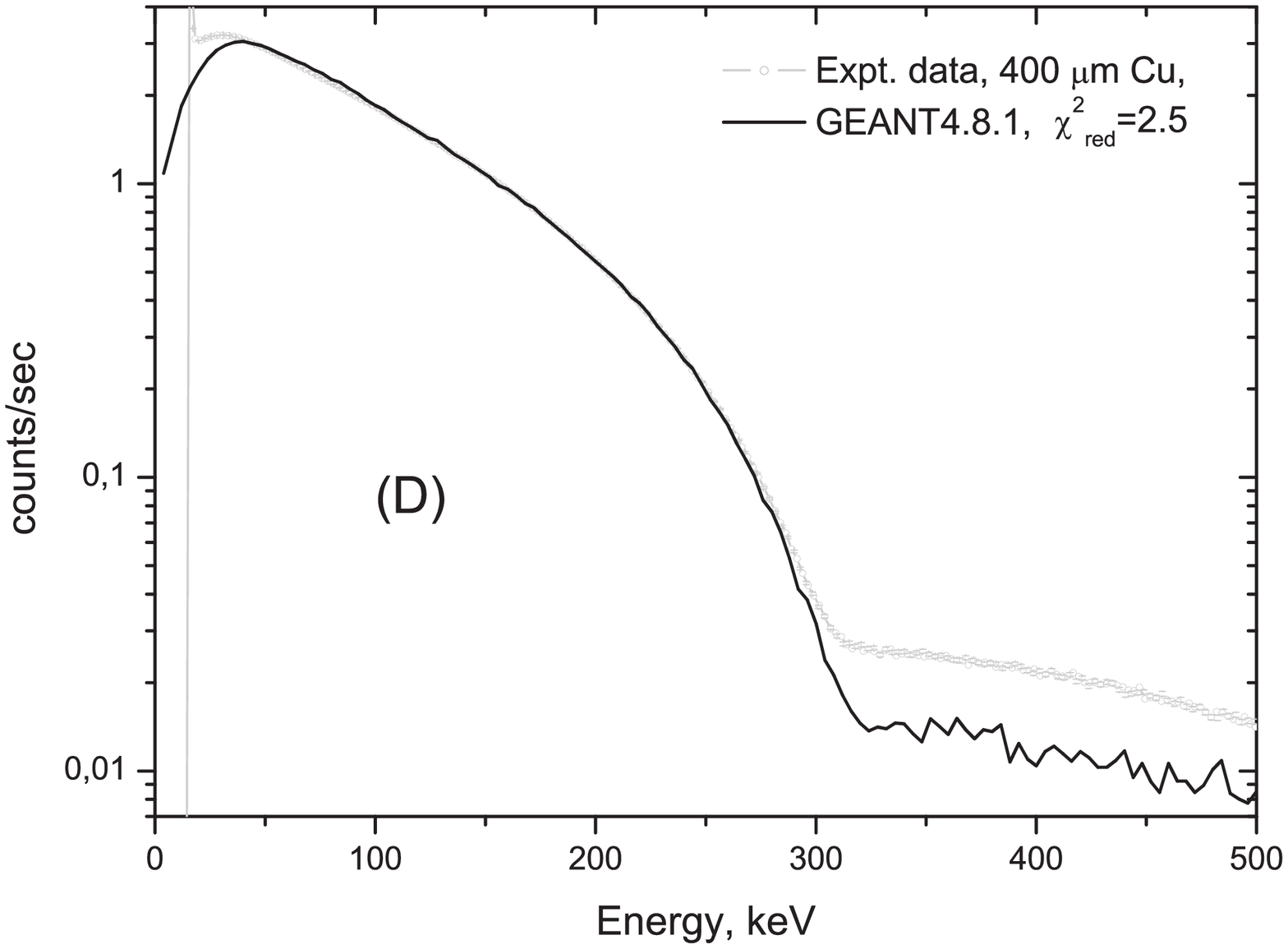}\\
\end{array}$
\end{center}
\caption{Experimental (panel A) and simulated (panel B) $^{60}$Co $\beta$ spectra for  the ``scattering-free'' source (see text) backed by copper foils with different thicknesses.
The two bottom panels compare experimental data and simulations for a Cu backing of 10 $\mu$m (panel C) and 400 $\mu$m (panel D). Note that for the spectrum obtained with a 400 $\mu$m backing, where the scattered fraction is largest, the reduced $\chi^2$ is slightly worse.}
\label{fig:backscat1}
\end{figure}

To estimate the influence of the source backing a series of measurements was performed. In these the $^{60}$Co $\beta$ spectrum was measured with copper foils of different thicknesses (i.e. 10, 20, 50, 100 and 400 $\mu$m) installed behind the scattering-free source. Simulations for these cases were then performed as well and compared to the experimental data. The results are shown in Fig. \ref{fig:backscat1}. The experimental data (panel ''a'') are normalized to the counting time. The simulated spectra shown in panel ''b'' are again rescaled to the corresponding experimental spectra by normalizing the integrals of the simulated and experimental spectra in the energy region from 150 keV to 300 keV, as mentioned above.

%In these simulations the pile-up probability was included too. This yielded a value of 0.11~\% (HOW WAS THIS DETERMINED ????????????), consistent with the value of XXXXXXXXXXX obtained by combining %the count rates and the shaping time.
%As can be seen pile-up influences mostly the region near the endpoint energy of $^{60}$Co and contributes to the Compton tail leading to better %agreement with the experimental data (see panels ''c'' and ''d'' of Fig. \ref{fig:backscat1}). The probability of pile-up was found by minimizing
%$\chi ^2$/$dof$ in the energy range 150 - 300 keV.

From the data shown in Fig.~\ref{fig:backscat1} the general effect of increasing the thickness of the backing is clearly visible: thicker backings cause a larger intensity in the low-energy part of the $\beta$ spectrum (caused by an increase in the backscattering from the source backing) as well as a larger contribution from Compton events. The backscattered fraction between 50 and 150 keV is clearly  saturating already between 50 and 400 $\mu$m, a feature that is reproduced by the simulations (see Figs.~\ref{fig:backscat1}(''a'') and ~\ref{fig:backscat1}(''b'')). A direct comparison of experimental data and simulations for Cu backings of 10 $\mu$m and 400 $\mu$m thickness is shown in Figs.~\ref{fig:backscat1}(''c'') and ~\ref{fig:backscat1}(''d''). It is seen that the simulations do very well between 50 and 300 keV for all backing thicknesses, in particular in the region between 50 and 150 keV which is directly sensitive to backscattering.

Above 300 keV (i.e. in the region that is dominated by Compton electrons) the data/simulation ratio is found to be about 1.5 (see e.g. Fig.~\ref{fig:BabyFr_Comparison_Difference.eps}a and Figs.~\ref{fig:backscat1}c and ~\ref{fig:backscat1}d), independent of backing thickness.  Simulations showed that for the setup in Fig.~\ref{fig:baby_fridge} that was used to measure the spectra shown in Figs.~\ref{fig:BabyFr_Comparison_Difference.eps} and \ref{fig:backscat1}, the source foil and the detector contributed about equally to the number of detected Compton electrons. The difference of about 50\% between the number of Compton electrons in the data and the simulations then suggests that either the amount of Compton scattering of $\gamma$ rays, or the description of multiple scattering of the Compton electrons, before they reach the detector, is not fully adequate (see also Ref.~\cite{Kadri:2007}).
%
%
%
%
%In panels ''c'' and ''d'' of Fig.~\ref{fig:backscat1} one notices that the simulated spectra underestimate the experimentally observed yield in the %energy region above the endpoint energy (by approx. a factor 1.5). The reason for this is still an open question. One of the explanations could be %the roughness (i.e. microscopic wrinkles) of the mylar foil on which the $^{60}$Co activity was dried. This could cause an additional scattering of %$\beta$ particles in the source, which was not taken into account in the simulations.  Since for the rescaling of the simulated spectra only the %energy region from 150 keV up to 300 keV was used, i.e. below the endpoint energy of $^{60}$Co, any small difference between simulations and %experiment in the description of the $\beta$ spectrum will then show up as a ''wrong'' normalization of the Compton tail.

However, this discrepancy between the data and simulations with respect to the number of Compton electrons will not have an important effect on the analysis of the $^{60}$Co $\beta$ anisotropy if the Compton background can be properly subtracted. In order to do so one has to know (1): the exact shape of the Compton background in the energy region of the $^{60}$Co $\beta$ spectrum and (2): the amplitude of this Compton background. As was shown above (see Fig. \ref{fig:Compton}) GEANT4 reliably describes the Compton background down to energies of about 100 keV, thus providing the shape of it. Further, knowing the shape of the Compton background its amplitude can be obtained from the part of the experimental spectrum above the $\beta$ endpoint. Applying this procedure to the data shown in Fig.~\ref{fig:BabyFr_Comparison_Difference.eps} causes the reduced $\chi^2$ to increase from 1.3 to 1.9. The subtraction is thus not perfect, and for analysis of real data an error related to this subtraction procedure will be included.

\section {Simulation of the $^{60}$Co spectrum in the real setup}

As a final step in the development of the simulation code that should take into account the effects that modify the experimental $\beta$ spectra, the $^{60}$Co $\beta$ spectrum was measured in the actual setup, this time also with non-zero magnetic field, and compared to simulations for these new and more delicate conditions. Three additional factors now had to be considered:
\begin{itemize}
  \item a realistic description of the $^{60}$Co source;
    \item the description of the propagation of the $\beta$ particles in the magnetic field;
    \item the anisotropy of the $\beta$ radiation emitted from a polarized sample.
\end{itemize}
These issues are considered in detail below.

\subsection{Description of the source}
\label{description of source}

The $^{60}$Co source was prepared by diffusing the $^{60}$Co activity into a 20 $\mu$m thick copper foil that was maintained at a temperature of 850$^\circ$C for 5 minutes. This procedure allowed to prepare a source with the radioactive $^{60}$Co nuclei being contained in an about 3~$\mu$m thin layer below the surface of the copper foil. The diffusion of cobalt in copper was studied in detail in Ref.~\cite{Mackliet1958}. Using the diffusion coefficient listed there for the temperature of 850 $^\circ$C at which our source was produced, we calculated the diffusion depth profile of the cobalt ions for this temperature and the diffusion time that was used. This profile was then implemented in the simulations.

\subsection{Description of the propagation of the $\beta$ particles in the magnetic field}
\label{propagation in field}

For the description of the magnetic field the field map in radial and axial directions with a step of 1 mm, provided by the manufacturer of the magnet (Oxford Instruments Ltd.), was implemented in the simulation code. In the GEANT4 toolkit Runge-Kutta integration is used to compute the motion of a charged particle in a general field \cite{webref:GEANT_MagneticField} and different methods of integration are provided. The specific method (or ''stepper'') can be chosen by the user, with the preferred choice being determined by the particular field conditions. In this work four different steppers provided by GEANT4 were compared: \textit{G4HelixExplicitEuler}, \textit{G4HelixImplicitEuler}, \textit{G4HelixSimpleRunge} and \textit{G4ClassicalRK4} (default stepper). The first three are specifically designed for the description of particle propagation in a purely magnetic field while the last one is a default GEANT4 stepper that can be used for general electromagnetic fields. The GEANT4.8.1 simulated spectra of $^{60}$Co obtained with the four different steppers were found to be very similar to each other. This is at variance with our earlier simulations performed with GEANT4.7.1 where the spectra simulated with different steppers were clearly differing. The best agreement with the experimental spectrum was then obtained for the \textit{G4HelixExplicitEuler}. This same stepper was also used in all later simulations with GEANT4.8.1.

\begin{figure}[h]
\begin{center}
$\begin{array}{c}
\epsfxsize=10cm
\epsffile{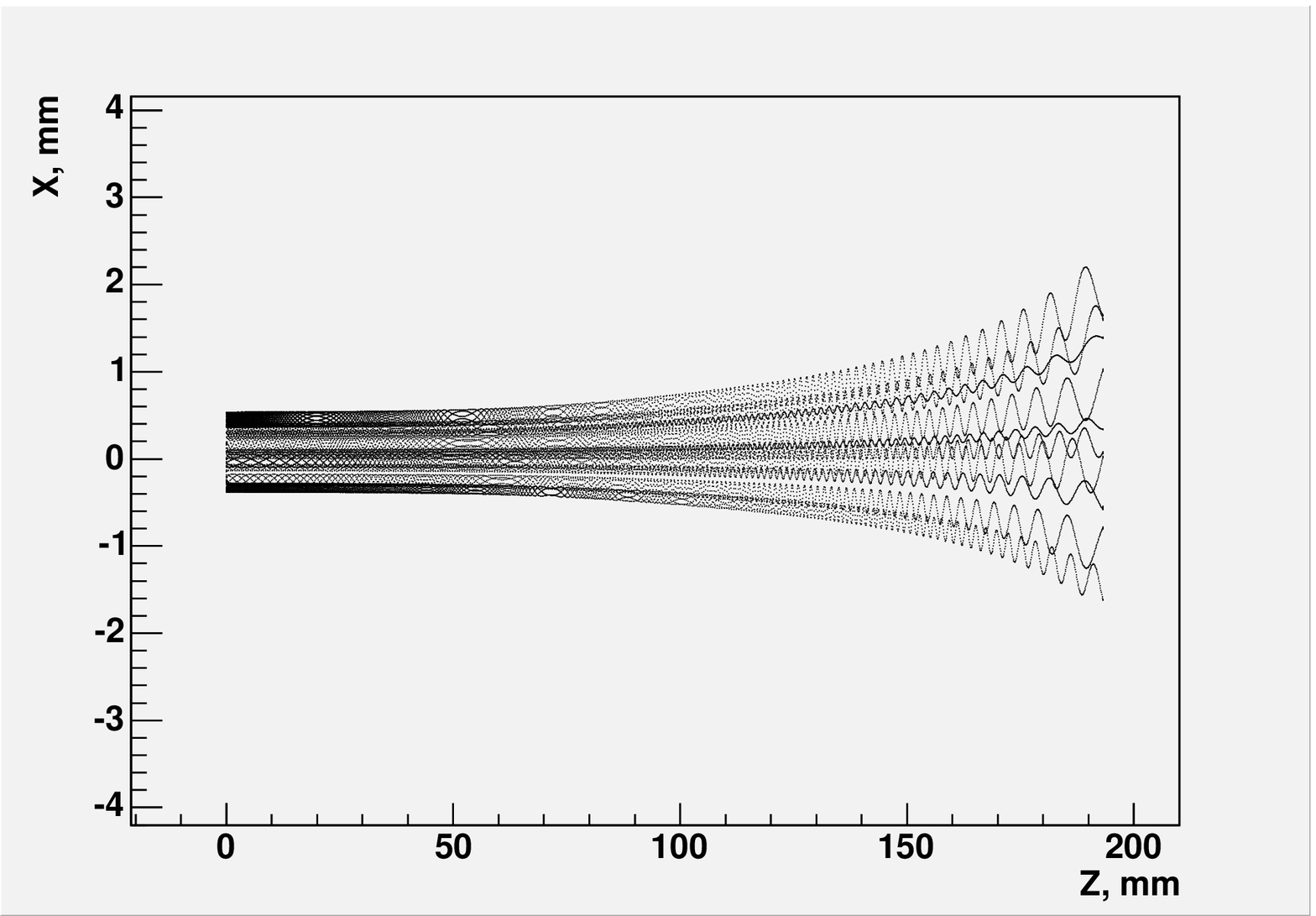}\\
\epsfxsize=10cm
\epsffile{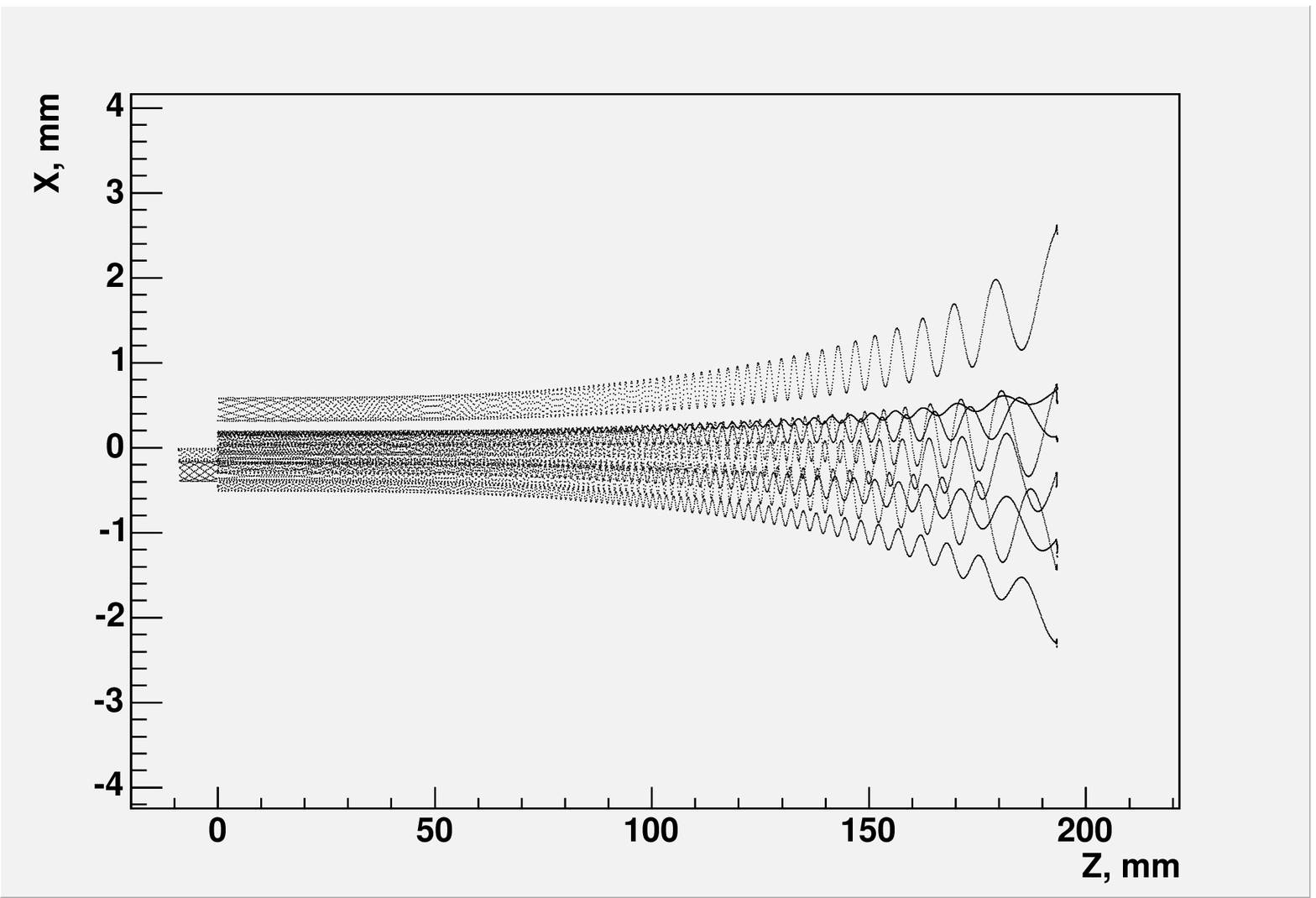}\\
\end{array}$
\end{center}
\caption{Trajectories of electrons with energies of 100 keV (upper panel) and 300 keV (bottom panel) in a magnetic field of 13 T. $X$-$Z$ plane; the magnetic field is along the $Z$-axis.}
\label{fig:trajXZ}
\end{figure}

Typical trajectories of electrons with energies of 100 keV and 300 keV in a 13 T magnetic field are shown in Fig \ref{fig:trajXZ}. The positions of the decaying $^{60}$Co nuclei emitting the electrons are chosen randomly in a circle with a diameter of 1 mm. The thickness of the copper host foil is 20 $\mu$m and the depth at which the decaying nuclei are sitting is randomized according to the depth distribution profile discussed above. %(Sect.~\ref{description of source}).

The different helix radii of electrons spiraling in the magnetic field are caused by the different energies and angles at which electrons leave the foil. Because of the very large magnetic field strength the electrons are strongly focused, leading to an effective detection solid angle of almost 2$\pi$. Further, due to the divergence of the magnetic field lines towards the detector and the principle of adiabatic invariance of the magnetic flux \cite{Jackson1975}, i.e. $p_\bot^2 / B$=const. (with $p_\bot$ the momentum component perpendicular to the magnetic field B), most of the transverse momentum of the electrons is transferred into longitudinal momentum. Therefore, no electrons arrive at the detector at angles of incidence with respect to the normal on the detector surface larger than about 20$^\circ$.

An interesting feature may finally be seen in the bottom panel of Fig. \ref{fig:trajXZ}, i.e. the backscattering of an electron on the sample holder (trajectory extending in the negative $Z$ direction). The sample holder has a 9 mm deep hole with a diameter of 6 mm and a rough surface (screw thread) behind the foil to avoid such backscattered events, but these can of course still occur from time to time on the bottom surface of this hole (see Fig.~\ref{fig:sketch}). Simulations showed, however, that this happens for only about one permille of the electrons that finally arrive in the detector.\\

\subsection{Anisotropy of the $\beta$ radiation}

To fully implement the GEANT4 based method described above for the analysis of experimental data, also the anisotropy of the $\beta$ radiation finally had to be included in the simulations. The anisotropy of the electrons emitted from $^{60}$Co was calculated using Eq. \ref{W} assuming the absence of any kind of exotic physics (i.e. assuming a pure $V-A$ interaction with maximal parity violation and no time-reversal violation). All terms in Eq.~\ref{W} are then known and the angular distribution of the $^{60}$Co $\beta$ decay electrons as a function of their emission angle $\theta$ can be implemented in the simulation code.

\begin{figure}[ht]
\begin{center}
$\begin{array}{c}
\mbox{\bf (a)}\\[-1.3cm]
\epsfxsize=11cm
\epsffile{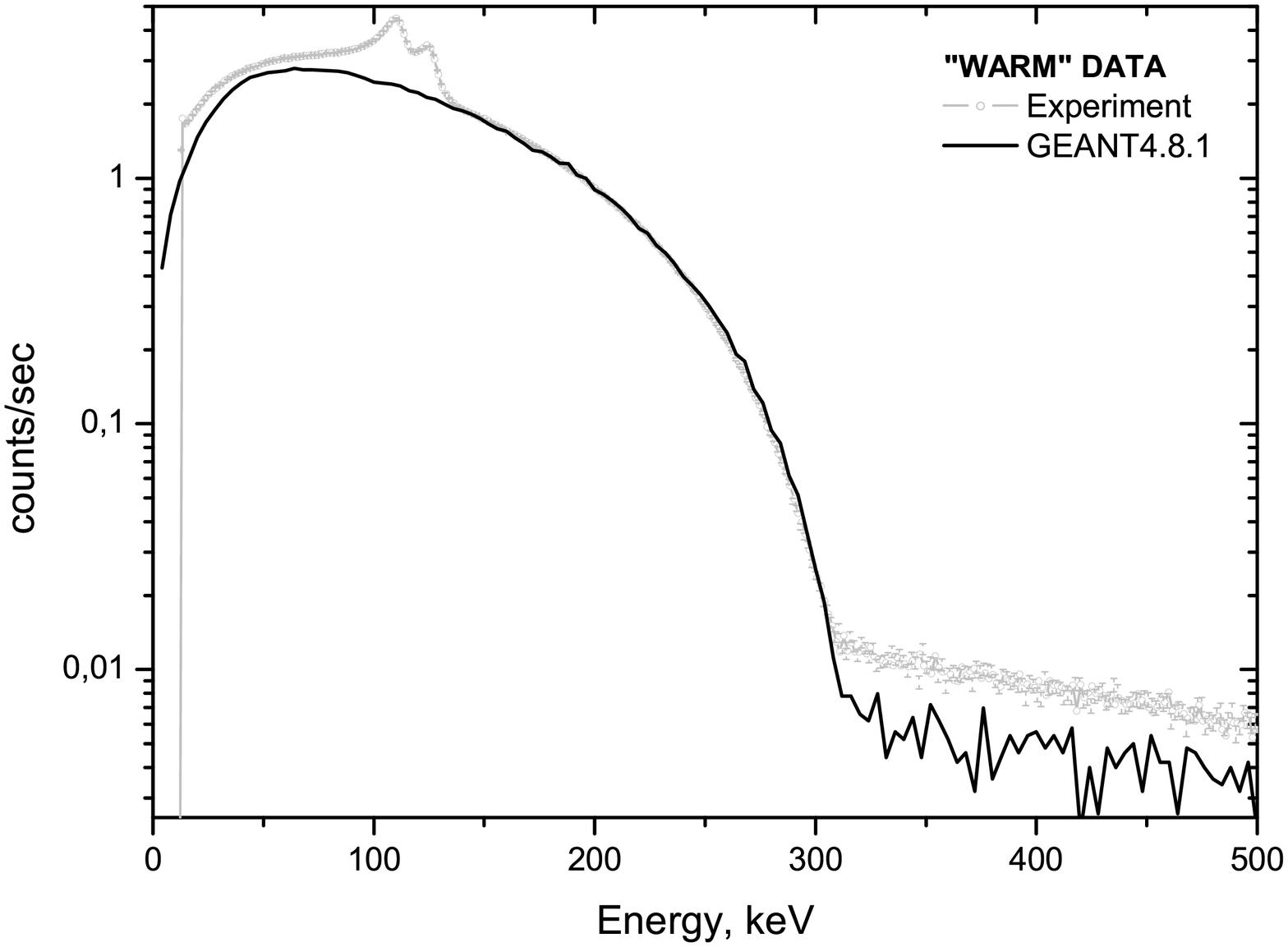}\\
\\%[0.5cm]
\mbox{\bf (b)} \\ [-1.3cm]
\epsfxsize=11cm
\epsffile{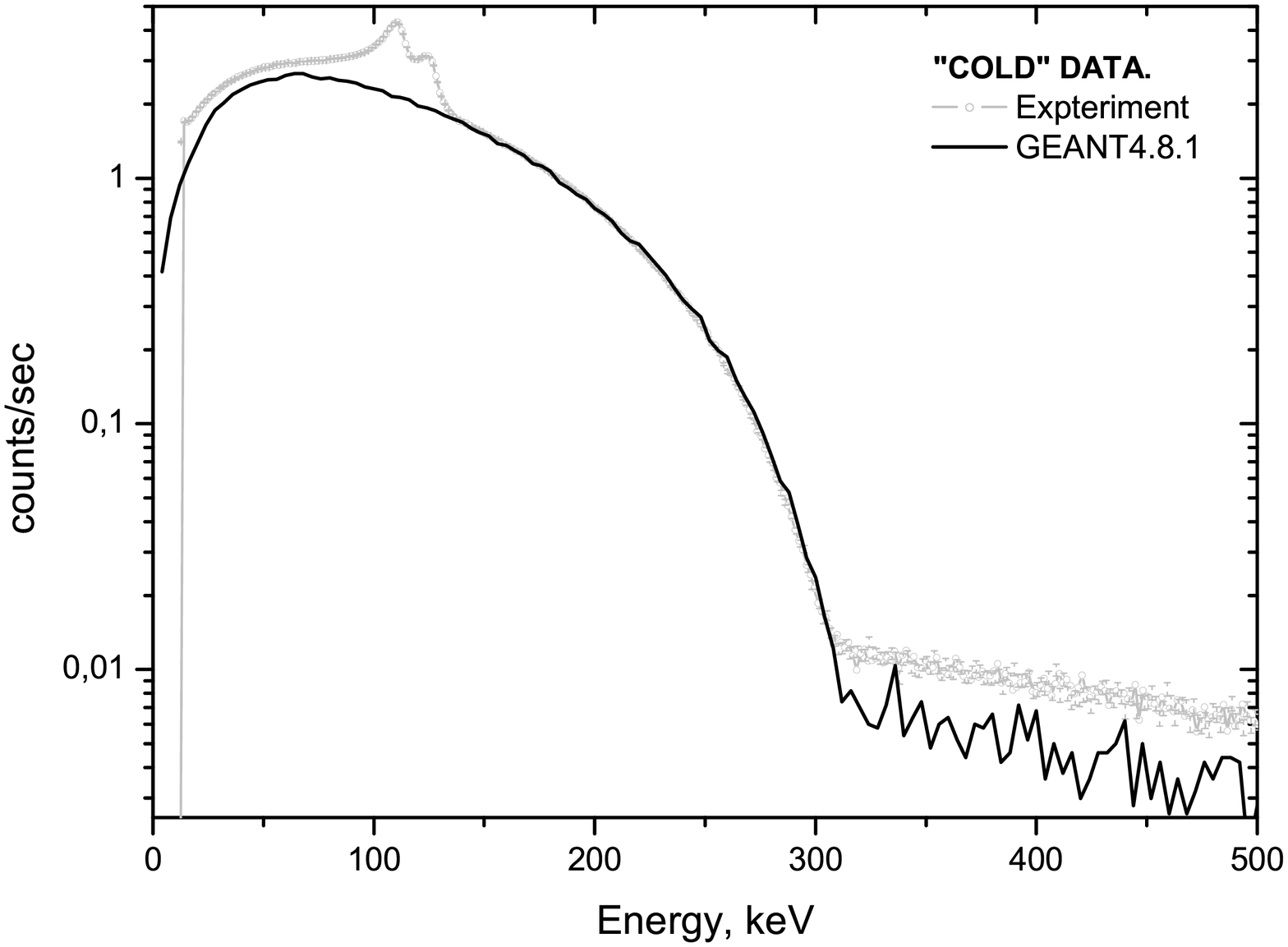}\\
\end{array}$
\end{center}
\caption{Comparison of the experimental and simulated $^{60}$Co spectrum for the actual experimental setup and a magnetic field $B$ = 13 T. Panel a: ``warm'' data (isotropic emission, unpolarized nuclei; $\chi^2/\nu$~=~1.2 for 40 degrees of freedom in the region of interest from 150 to 300 keV); panel b: ``cold'' data (anisotropic emission, polarized nuclei; $\chi^2/\nu$~=~0.8 for 40 degrees of freedom in the region from 150 to 300 keV). The peaks between 100 keV and 150 keV in the experimental $\beta$ spectrum are conversion electrons from $^{57}$Co. The Compton tail in the experimental spectra also includes a small amount of Compton background from the $^{54}$Mn thermometer source that was also present on the sample holder. Note that the $^{57}$Co and $^{54}$Mn sources were not included in the simulations.}
\label{fig:G481_warmANDcold}
\end{figure}

In Fig.~\ref{fig:G481_warmANDcold} the simulated isotropic (i.e. unpolarized nuclei, panel ``a'') and anisotropic (polarized nuclei, panel ``b'') spectra are compared with experimental data obtained with $^{60}$Co. For the experimental spectra a pile-up probability of 0.18~\% was obtained from the shaping time and count rate. This value was then used in the simulations. In the analysis of future measurements we will assign a generous error (i.e. of ~50~\%) to this estimated pile-up rate. Further, as we are always measuring at not too high counting rates (typically about 1 kHz or lower) the pile-up correction will not induce a large effect in the simulated spectra and will therefore not have a significant effect on the final result.

In Ref.~\cite{Kadri:2007} it was shown that simulation results do not change anymore when values for the cut-for-secondaries parameter $CFS$ smaller than 1~$\mu$m are used. For such small values simulation times become exceedingly large, however. Simulations were therefore performed for $CFS$ = 1~$\mu$m and 10~$\mu$m. As no difference could be observed within the statistical errors of these simulations the value of the $CFS$ parameter was set to 10 $\mu$m so as to obtain reasonable simulation times.

The experimental spectra in Fig.~\ref{fig:G481_warmANDcold} are ``distorted''
in the energy region below 150 keV by conversion electrons from the $^{57}$Co activity
that was diffused in the same foil as $^{60}$Co, for thermometry. Good
agreement between experimental data and simulations is again obtained in
the energy region from 150 to 300 keV, in spite of the fact that the
amplitude of the Compton tail above the spectrum endpoint is again too
low in the simulated spectra. In Fig.~\ref{fig:backscat1} it was shown
already that the Compton fraction depends on the source conditions
(as e.g. the thickness of the backing), but there may be more parameters
determining this. Simulations showed that for the setup with
the 13 T magnetic field that was used here about 99~\% of the detected Compton
electrons originate from the Cu source foil (compared to an about equal number
of Compton electrons being produced in the source foil and in the detector
when no magnetic field is present; cf. end of section 4). This can be
easily understood. A Compton electron created in the source will be
focussed directly on to the detector by the strong magnetic field, similar to
the $\beta$ decay electrons. However, in order to have Compton scattering
occur in the detector, $\gamma$ rays from the sample first have to reach the
detector, which has a very small (geometrical) solid angle (order 10$^{-4}$).
The remaining 1~\% of the detected Compton electrons mainly originate from
the Cu sample holder. Compton electrons created elsewhere in the setup were
found to have a negligibly small probability to make it to the detector.

Simulations also showed that the presence of the weak $^{54}$Mn thermometer
(with a 835 keV $\gamma$ ray) caused a negligible amount of Compton
electrons reaching the detector. This can be easily understood from the
geometry of the setup as shown in Fig.~\ref{fig:sketch}: most Compton
electrons from gamma rays of the $^{54}$Mn source are indeed absorbed in the part of
the sample holder that is situated between the $^{54}$Mn source and the Si PIN diode
detector.

The $^{57}$Co conversion electron peaks could be well reproduced in the simulations.
But to make these cause a significant increase in the amount of counts above the spectrum
endpoint in the simulations, a much too large pile-up percentage had to be assumed (i.e. of
the order of 1~\% or more, compared to an expected pile-up rate of about 0.2~\% given the count rates and
amplifier shaping times). One is therefore again lead to conclude that the counts
above the beta spectrum endpoint most probably are Compton electrons related to the
gamma rays of $^{60}$Co. Then, either the amount of Compton scattering
of $\gamma$ rays in the backing, or the description of multiple scattering
so that they can reach the detector, is not fully adequate.
It was found that when assuming the activity to be
distributed throughout the entire foil (which would have required a much longer
diffusion time than was actually used), leading to an increased amount of Compton
scattering in the backing, the amount of Compton counts in the spectrum relative
to the number of detected $\beta$ particles increases only by about 20~\%. This
is less than half of the 50~\% increase that is needed. This could indicate that it is
the description of multiple scattering of the Compton electrons, so that they can
reach the detector, that is not fully adequate. It is not clear yet why the simulation
code does not describe this correctly and of course this will have to be taken
into account as a systematic error in the analysis of experimental data with this code.

\section{Conclusion}

A GEANT4 based Monte-Carlo routine was developed and optimized to simulate the spectrum shape for $\beta$ particles emitted by unpolarized as well as polarized nuclei in LTNO experiments and to extract the $\beta$ asymmetry parameter using Eq. \ref{eq:ratio}. The code accounts for the different effects that modify the emitted $\beta$ spectrum, i.e. (back)scattering, magnetic field effects, etc.\\
Good agreement between simulations and experimental data in the energy region of interest is found for a 500 $\mu$m Si PIN diode detector with the relative difference always being less than about 2\%. Similar results were obtained by other authors \cite{Martin:2003}-\cite{Golovko:2008}. %\cite{Martin:2003,Martin:2006,Kadri:2007,Golovko:2008}.

Due to the modularity of this simulation code and the
object-oriented structure of the GEANT4 toolkit this method can
be easily adopted for similar types of measurements with a
different geometry and a different isotope studied. In particular it is now
being implemented for a series of precision measurements of the
$\beta$ asymmetry parameter of selected isotopes that is being
performed both in Leuven and
at the NICOLE low temperature nuclear orientation setup at
ISOLDE/CERN \cite{Wauters2009,Severijns2004}. The code can in
principle rather easily be generalized to other be used also
for experimental setups using other polarization techniques.

\end{document}